\documentclass[3p,twocolumn]{elsarticle}
\usepackage{graphicx}
\usepackage{amsmath,amssymb,physics,dsfont}
\usepackage{mathrsfs}
\usepackage{color}
\usepackage{defs-private}
\usepackage{microtype}
\usepackage{braket}
\usepackage{tabularray}

\begin{document}
\begin{frontmatter}
\title{exaPD: A highly parallelizable workflow for multi-element phase diagram (PD) construction}
\author[a,b]{Feng Zhang\corref{cor1}}
\ead{fzhang@ameslab.gov}
\author[a]{Zhuo Ye}
\author[c]{Maxim Moraru}
\author[c]{Ying Wai Li}
\author[a]{Weiyi Xia}
\author[a,b]{Yongxin Yao}
\author[a,b]{Cai-Zhuang Wang}

\affiliation[a]{organization={Ames National Laboratory, U.S. Department of Energy}, addressline={Ames, Iowa 50011}, country={USA}}
\affiliation[b]{organization={Department of Physics and Astronomy, Iowa State University}, addressline={Ames, Iowa 50011}, country={USA}}
\affiliation[c]{organization={Los Alamos National Laboratory}, addressline={Los Alamos, NM 87545}, country={USA}}

\cortext[cor1]{Corresponding author}

\begin{abstract}
Phase diagrams (PDs) illustrate the relative stability of competing phases under varying conditions, serving as critical tools for synthesizing complex materials. Reliable phase diagrams rely on precise free energy calculations, which are computationally intensive. We introduce exaPD, a user-friendly workflow that enables simultaneous sampling of multiple phases across a fine mesh of temperature and composition for free energy calculations. The package employs standard molecular dynamics (MD) and Monte Carlo (MC) sampling techniques, as implemented in the LAMMPS package. Various interatomic potentials are supported, including the neural network potentials with near {\it ab initio} accuracy. A global controller, built with Parsl, manages the MD/MC jobs to achieve massive parallelization with near ideal scalability. The resulting free energies of both liquid and solid phases, including solid solutions, are integrated into CALPHAD modeling using the PYCALPHAD package for constructing the phase diagram.      
\end{abstract}

\end{frontmatter}
\section{Introduction}
Computational materials discovery has advanced rapidly, driven by enhanced computational power, and AI/ML techniques~\cite{Gubernatis2021,Vasudevan2021,Kusne2014,Kabiraj2020,cai2020,Katsikas2021,Rhone2020,Landrum2003,Merchant2023,Szymanski2023,Mroz2022}. However, only a small fraction of the predicted materials have been experimentally validated due to limited knowledge of viable synthetic pathways~\cite{Aykol2019,Chen2024}. Reliable multi-element phase diagrams are essential for resolving the thermodynamic competition among relevant phases under synthetic conditions, and thus are important for predicting synthesizability and suggesting synthetic pathways. Constructing these phase diagrams computationally requires highly accurate free-energy calculations. {\it Ab initio} methods, such as the density-functional theory (DFT), provide reliable energetics at 0 K. However, due to its limitations on length and time scales, it is difficult to address many finite-temperature effects such as the anharmonicity in solids and the amorphicity of liquids. Classical force fields, while computationally efficient, often lack the quantum-mechanical accuracy needed for complex materials. 

Recent breakthroughs in artificial neural network potentials (NNP) have addressed these challenges~\cite{Behler2017,Blank2019,Zhang2018}. NNPs maintain near ab initio accuracy while extending the length and time scales to thousands of atoms and nanoseconds, respectively, making it feasible to implement many accurate methods for free energy calculations. These methods include the thermodynamic integration (TI) for solid and liquid phases~\cite{Kirkwood1935,Frenkel2023}, the solid-liquid coexistence method for measuring the melting point~\cite{Morris1994}, and several different flavors of hybrid Molecular Dynamics (MD) and Monte Carlo (MC) methods for solid solutions~\cite{Sadigh2012,Li2024}. 

Despite these advances, it remains computationally intensive to construct a multi-element phase diagram. Thousands of MD or MC jobs are required to sample competing phases on a fine mesh of state parameters, including temperature, pressure, composition, etc. Fortunately, the exascale computing era offers unprecedented computational power and capablilities. With minimal communications and dependencies between jobs, high scalability would be achievable on exascale machines. We introduce exaPD, a workflow that orchestrates all necessary jobs for multi-element phase diagram construction using LAMMPS, a mature and flexible package for atomistic simulations~\cite{Plimpton1995,Thompson2022}.  A global controller powered by Parsl, an open-source package for parallel programming in Python~\cite{Babuji2019}, ensures efficient job management on high-performance computers, achieving near-ideal scalability. The resulting free energies are post-processed with CALPHAD modeling. Unlike similar packages~\cite{Menon2021}, exaPD prioritizes maximal parallelization and scalability. 

The paper is organized as follows. We begin with an overview of the methods for free energy calculations employed in the workflow, including benchmarking examples. We then introduce the global controller, followed by a detailed description of the main components of the workflow and its JSON-based user interface. Finally, we conclude with a summary and future outlook.


\section{Computational methods}
The general workflow is developed within the framework of the thermodynamic integration (TI)~\cite{Kirkwood1935,Frenkel2023}, which is based on the fact that a derivative of the free energy with respect to a state parameter can usually be expressed as the ensemble average of a quantity that is readily measurable in a single molecular dynamics (MD) or Monte Carlo (MC) simulation. Then, the free energy difference between the initial and final states is obtained by integrating the derivative along a reversible path. In practice, one can independently sample a series of well-equilibrated data points along the integration and perform the integration numerically. Alternatively, one can use nonequilibrium sampling techniques~\cite{Konning2025}, in which the corresponding state parameter is allowed to switch continuously from the initial state to the final state and back to the initial state again in a single simulation, so that the energy dissipation due to the finite switching time can be largely canceled out in the forward and backward processes. We will follow the strategy of equilibrium sampling to maximize parallelizability in our approach. 

Traditionally, the state parameter $\lambda$ is introduced to create an artificial intermediate state with a Hamiltonian between the real state and a reference state whose free energy is already known: $\h_\lambda
=(1-\lambda)\h_R+\lambda \h$, where $\h$ and $\h_R$ are the Hamiltonian of the true physical system and the reference system, respectively. The Helmholtz free energy difference between the two systems can be evaluated as 
\begin{equation}\label{eq:ti}
    F-F_R=\int_0^1\langle \h-\h_R\rangle_{\lambda,NVT}~d\lambda,
\end{equation}
where $\langle\cdots\rangle_{\lambda,NVT}$ denotes the canonical ensemble ($NVT$) average with respect to the intermediate Hamiltonian $\h_\lambda$. Since the kinetic energy contribution to the Hamiltonian is the same for the real and reference systems, $\langle \h-\h_R\rangle$ amounts to the potential energy difference $\langle U-U_R\rangle$.
In different variations, the state parameter used in TI can also be the temperature ($T$), the pressure ($P$), or the composition ($x$). In the following, we briefly describe the process of calculating the free energy of both solid and liquid phases, together with validating and benchmarking examples.

\subsection{Free energy of line compounds}~\label{subsec:linecomp}
We start with ordered stoichiometric phases that appear as a vertical line in phase diagrams and thus sometimes are referred to as line compounds. Because it is completely ordered, the configurational entropy plays no role in this phase. It has been well established that the Einstein crystal is a suitable reference system for the free energy calculation of this type of compounds~\cite{Frenkel1984}. In an Einstein crystal, all atoms are simple harmonic oscillators bound to their equilibrium positions, and its free energy is expresses as $F_E=3Nk_\textrm{B}T\sum_\alpha x_\alpha\ln(\hbar\omega_\alpha/k_\textrm{B}T)$, where $N$ is the number of atoms, $x_\alpha$ the composition for the $\alpha$ element, and $\omega_\alpha$ the angular frequency of the harmonic oscillator for the $\alpha$ element. Generally, $\omega_\alpha$ does not need fine-tuning, and an estimation based on the phonon spectrum of the real system will be sufficient. A common approach is to set the spring constant $k_\alpha=3k_\textrm{B}T/\langle\Delta r_\alpha^2\rangle$, so that the Einstein crystal and the real system have the matching mean-square displacement ($\langle\Delta r^2\rangle$) for each element~\cite{Frenkel2023,Freitas2016}. $\omega_\alpha$ can be calculated as $\omega_\alpha=\sqrt{k_\alpha/m_\alpha}$ ($m_\alpha$ is the atomic mass for element $\alpha$).

To perform the TI, one first thermalizes a supercell of the target crystalline phase with a cubic-like shape at the target temperature and pressure under the isothermal-isobaric ensemble ($NPT$) to obtain the equilibrium volume. The mean-square displacement of each element is also measured in this process for the determination of the spring constants. Then the supercell is quenched to 0 K with the volume fixed to bring all atoms to the equilibrium positions for applying the spring forces. MD jobs are set up for a series of equidistant $\lambda$ values between 0 and 1 with the default $\Delta\lambda=0.05$ to sample $\langle U-U_R\rangle_{\lambda,NVT}$ for the intermediate Hamiltonian $\h_\lambda$, based on which the integration in Eq.~\ref{eq:ti} can be evaluated to obtain the Helmholtz free energy $F$. Since $F$ is evaluated at the equilibrium volume under the pressure $P$, the Gibbs free energy $G$ under this pressure is simply $G=F+PV$. 

While in principle one can repeat the above process to calculate the Gibbs free energy at other temperatures, a more efficient process is to use the Gibbs-Helmholtz equation
\begin{equation}\label{eq:gibbs-helmoltz}
    \frac{G(T,P)}{T}-\frac{G(T_0,P)}{T_0}=\int_{T_0}^{T}-\frac{H(\T,P)}{\T^2}d\T,
\end{equation}
where $T_0$ is the temperature at which the TI is implemented, and $H(T,P)$ is the enthalpy of the system. In practice, one samples $H(T,P)$ at a few discrete temperatures (the default $\Delta T$ is 50 K) that allow a smooth interpolation between $T_0$ and an ending temperature; then $G(T,P)$ can be calculated at an arbitrary temperature according to Eq.~\ref{eq:gibbs-helmoltz}. 

\begin{figure}[tb]
    \centering
    \includegraphics[width=\linewidth]{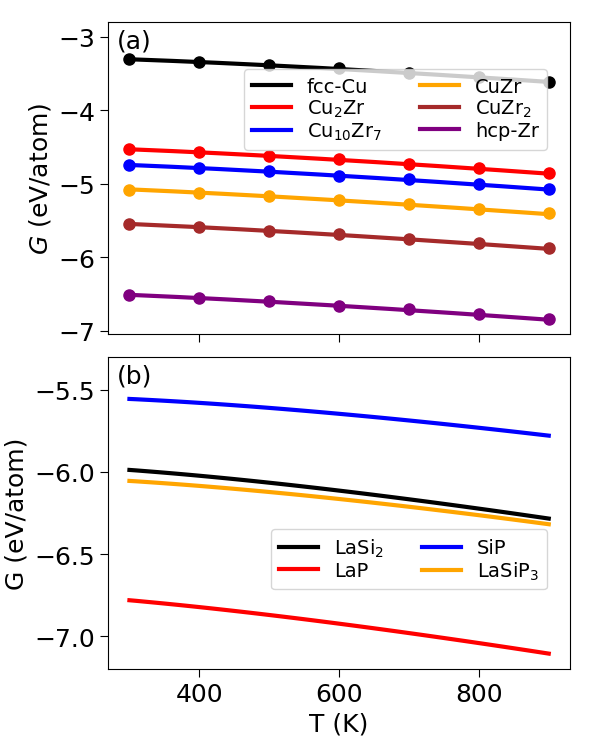}
    \caption{Gibbs free energy as a function of the temperature for various compoinds in (a) the Cu-Zr system; and (b) the La-Si-P system. EAM-FS and NNP potentials are used for the Cu-Zr and La-Si-P system, respectively. The solid circles in (a) are results from Ref.~\cite{Tang2012} for comparison.}
    \label{fig:solid-einstein}
\end{figure}

To validate our approach, we first repeat the calculation of the Gibbs free energy for various compounds in the Cu-Zr system, using a widely used embedded-atom model (EAM) potential in the Finnis-Sinclair (FS) format~\cite{Daw1984,Finnis1984,Mendelev2009}. Fig.~\ref{fig:solid-einstein} (a) shows the Gibbs free energy as a function of temperature at the ambient pressure for fcc-Cu, hcp-Zr, and several intermetallic compounds. The solid circles are results from Ref.~\cite{Tang2012}, in which TI was performed separately for each temperature. One can see excellent agreement between these two approaches. In Fig.~\ref{fig:solid-einstein} (b), we present the free energy of compounds in the La-Si-P system, using a newly developed NNP~\cite{Tang2025}. Structures from each sub-binary system and a ternary $\textrm{LaSiP}_2$ phase were tested, confirming the workflow's compatibility with ternary systems and more accurate NNPs. 


\begin{figure}[tb]
    \centering
    \includegraphics[width=\linewidth]{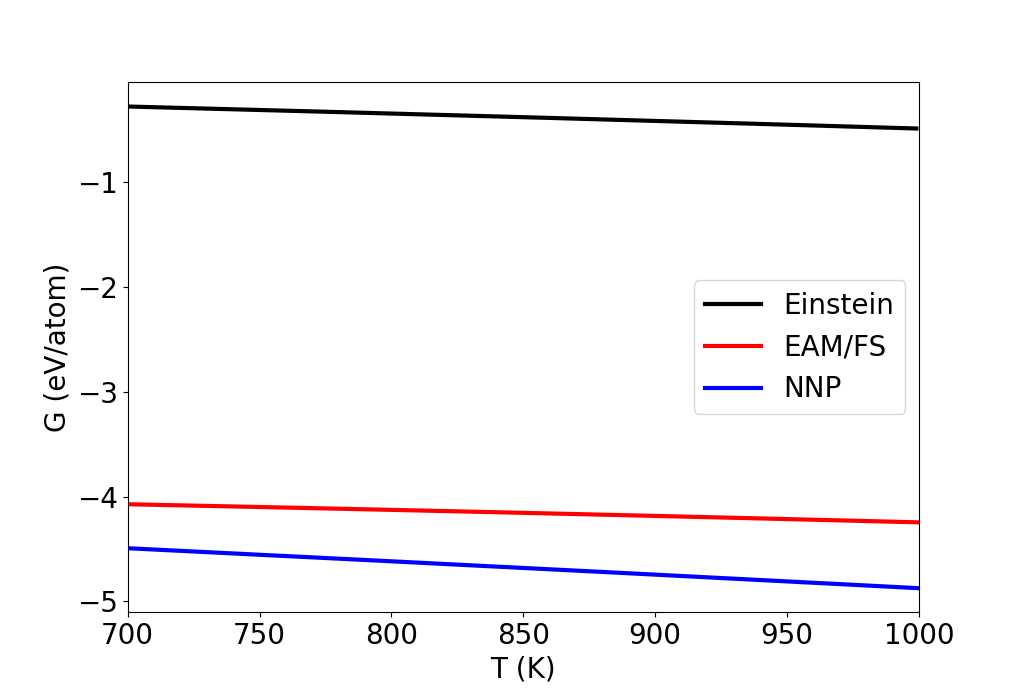}
    \caption{Gibbs free energy of fcc-Al as a function of the temperature calculated using the Einstein model, an EAM-FS potential, or a neural-network trained potential. The Einstein crystal is used as the reference for the EAM-FS potential in the thermodynamic integration; while the EAM-FS potential is used as the reference for the NNP. }
    \label{fig:fcc-al}
\end{figure}

In addition to using the Einstein crystal as a reference system, thermodynamic integration (TI) can facilitate a transformation between different interatomic potentials. This approach is advantageous when the initial auxiliary potential is computationally efficient, enabling high accuracy at a low cost, while the target potential is more expensive. By starting from an auxiliary potential closer to the target potential than the traditional Einstein crystal, significant computational savings can be achieved~\cite{Menon2021,Sun2023}. Fig.~\ref{fig:fcc-al} displays the Gibbs free energy as a function of the temperature for fcc-Al, computed using an Einstein crystal with the spring constant $k=2.6$ eV/\AA$^2$, an EAM-FS potential~\cite{Mendelev2015}, and a NNP~\cite{Tang2020}. The Einstein crystal is used as the reference for the EAM-FS potential, which in turn acts as the reference for the NNP. The free energy difference between EAM-FS and NNP is significantly smaller than that between EAM-FS and the Einstein crystal, demonstrating the efficiency of this approach.

\subsection{Free energy of solid solutions}~\label{subsec:sgmc}
\label{subsec:linecompount}

Another important type of solid phase is the disordered solid solutions, in which the configurational entropy makes a non-negligible contribution to the free energy. It is difficult to sample the configurational space in conventional MD due to its limitations on length and time scales. Hybrid MC/MD techniques of different variations have been proposed to circumvent this problem. In these methods, atoms are allowed to swap and/or transmute in addition to following the trajectories governed by the Newtonian equations of motion. We implement a semi-grand canonical ensemble (SGCE) technique~\cite{Sadigh2012}, in which the total number of the atoms in the simulation cell is fixed while the composition can change depending on the chemical potential difference ($\Delta\mu$). Taking a binary system $\textrm{A}_{1-x}\textrm{B}_x$ as an example, $\Delta\mu\equiv\mu_\textrm{B}-\mu_\textrm{A}=\frac{\partial G}{\partial x}$. After a certain number of regular MD steps, a randomly selected atom is tried to change its type according to the Metropolis principle, that is, the acceptance rate $r=\min(1,e^{-(\Delta U+\Delta\mu N\Delta x)/k_\textrm{B}T})$, where $\Delta U$ and $\Delta x$ are the change in the total potential energy and the composition after the transmute, respectively. The above process is repeated until an equilibrium is established. In this way, one can sample the relation between the equilibrium composition $x$ and the chemical potential difference $\Delta\mu$. By integrating the function $\Delta\mu(x)$, one can obtain the Gibbs free energy difference between the disordered alloy $\textrm{A}_{1-x}\textrm{B}_x$ and the end members pure A or B, whose free energy can be calculated using the method described in the previous section. 

\begin{figure}[tb]
    \centering
    \includegraphics[width=0.9\linewidth]{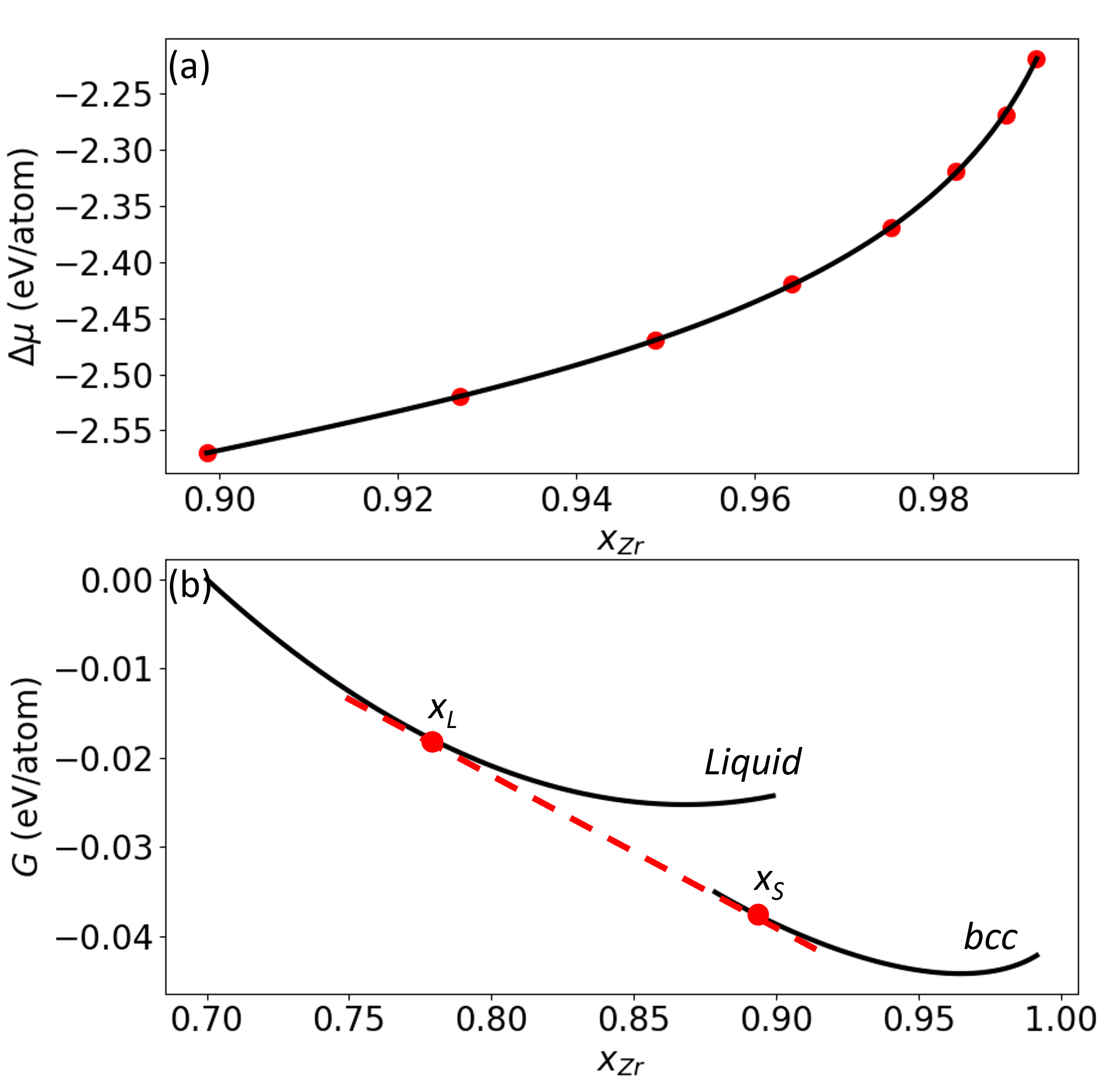}
    \caption{Semi-grand canonical calculation of the bcc phase in the Cu-Zr system. (a) The chemical potential difference as a function of the Zr composition at T = 1600 K. The red circles are from the MD simulations, and the solid line is a fitting to the derivative of the RK polynomial. (b) The calculated Gibbs free energy of both the liquid and bcc phase as a function of the Zr composition at T = 1600 K. The pure Zr and Cu$_{70}$Zr$_{30}$ liquids are used as the reference states. The dashed red line is a common tangent construction, which gives the Zr compositions in the liquid and bcc phases. }
    \label{fig:sgmc-bcc}
\end{figure}

In CALPHAD modeling, the molar Gibbs free energy of a binary non-ideal solution phase is usually represented by the following equation
\begin{multline}\label{eq:sol}
    G(x,T) = (1-x)G_0(T) + xG_1(T) + RT[x\ln x  \\
    + (1-x)\ln(1-x)] + x(1-x)\Omega(x,T),
\end{multline}
where $G_0(T)$ and $G_1(T)$ are the molar Gibbs free energy for the two end members corresponding to $x=0$ and $x=1$, respectively, $R$ is the gas constant, and $\Omega(x,T)$ is the Redlich-Kistler polynomial expressed in powers of $(1-x)-x=1-2x$:
\begin{multline}\label{eq:rk}
    \Omega(x,T)=L_0(T)+L_1(T)(1-2x)+L_2(T)(1-2x)^2 \\
    +L_3(T)(1-2x)^3.
\end{multline}
Here, we keep terms up to the third power of $1-2x$. Instead of directly integrating $\Delta\mu(x)$, we fit $\Delta\mu(x)$ to the derivative $\frac{\partial G}{\partial x}$ that can be readily calculated from Eqs.~\ref{eq:sol} and~\ref{eq:rk}:
\begin{multline}\label{eq:sol-der}
    \Delta\mu(x,T)=G_1(T)-G_0(T)+RT\ln\frac{x}{1-x} \\
    +(1-2x)\Omega(x,T)+x(1-x)\frac{\partial\Omega}{\partial x}.
\end{multline}
There are 4 fitting parameters, $L_0$ to $L_3$, as given in Eq.~\ref{eq:rk}. This process is repeated at several different temperatures to capture the temperature-dependence of the Redlich-Kistler polynomial. 

As an example, we calculated the Gibbs free energy for a Zr-rich bcc phase $\textrm{Cu}_{1-x}\textrm{Zr}_x$. A newly developed EAM-FS potential is used in this calculation~\cite{Mendelev2019}. We start by calculating the free energy of the pure bcc-Zr phase ($x=1$) using the Einstein crystal as the reference state. Then, SGCE simulations are carried out across the temperature range $1000~\textrm{K}\leq T\leq2000~\textrm{K}$ with the increment $\Delta T=100$ K. A series of chemical potential differences is implemented at each temperature, assuring that the solid phase remains stable at the largest $\Delta\mu$ magnitude without melting. We use $T=1600$ K as an example to show how to determine the composition of the liquid and solid phases in equilibrium. Fig.~\ref{fig:sgmc-bcc} (a) shows $\Delta\mu$ as a function of Zr composition at $T=1600$ K, with the red circles denoting the raw measurements in MD simulations and the solid line the fitting to Eq.~\ref{eq:sol-der}. In Fig.~\ref{fig:sgmc-bcc} (b), we show the calculated Gibbs free energy of the bcc solid solution phase together with that of the liquid phase at the same temperature of 1600 K. To better reveal the non-linear nature of the composition dependence, the Gibbs free energy for both the solid and liquid phases is referenced to the liquid phase with $x=0$ and $x=0.7$. The dashed red line is a common tangent, showing the composition of the solid and liquid phases to be $x_L=0.78$ and $x_S=0.89$, respectively. This determination of the transition compositions with the common-tangent construction is not affected by the reference states.

\subsection{Free energy of liquids}~\label{subsec:liquid}
The TI technique is also widely used for liquid free energy calculations. A natural reference system is the non-interacting ideal gas whose exact free energy is readily available. However, in many cases, the thermodynamic path needs to be carefully constructed to avoid a liquid-vapor phase transition~\cite{Abramo2015}. Numerical issues can also arise when the system approaches the ideal gas in the low-density or weak-interaction limit, causing relatively large errors. An alternative choice as the reference system is the Uhlenbeck-Ford model (UFM)~\cite{Leite2016}, whose potential energy is defined as:
\begin{equation}\label{ufm}
    U_\textrm{UF}(r)=-\frac{p}{\beta}\ln\left(1-e^{-(r/\sigma^2)}\right),
\end{equation}
where $\beta=1/k_\textrm{B}T$, $p$ is a dimensionless scaling factor for the interaction strength, and $\sigma$ is a scaling factor for the distance. The UFM is purely repulsive and maintains a single stable liquid phase under all conditions. Thus, by using the UFM as the reference system, one effectively eliminates possible hysteresis from phase transitions. The equation of state of the UFM at the low-density limit can be derived analytically, while at normal densities, it can be reliably obtained through atomic simulations~\cite{Leite2016}. Consequently, the excess free energy of the UFM, defined as the free energy difference relative to the ideal gas, has been accurately determined for several values of $p$~\cite{Leite2016}.

The UFM can be used as the reference for both pure and alloy systems~\cite{Leite2019}. Alternatively, we have introduced an alchemical TI method for calculating the free energy of a liquid alloy $\textrm{A}_{1-x}\textrm{B}_x$, using the pure $\textrm{A}$ liquid as the reference~\cite{Yang2018}. Here, we use a binary liquid to illustrate this approach. The workflow supports a general multi-element system. The same method was implemented in the package calphy for free-energy calculations~\cite{Menon2021}. As chemically different as elements A and B can be, the $\textrm{A}_{1-x}\textrm{B}_x$ alloy should still be much closer to the A system than to the purely repulsive UFM. In this alchemical approach, one first uses the standard TI technique as described in Eq.~\ref{eq:ti} to transfer the $\textrm{A}_{1-x}\textrm{B}_x$ liquid to the $\textrm{A}_{1-x}\textrm{B}^\prime_x$ liquid, in which the factitious $\textrm{B}^\prime$ atom has the same mass as $\textrm{B}$ but interacts in exactly the same way as A. The $NPT$ ensemble can be used to directly calculate the Gibbs free energy difference. Then, the mass of $\textrm{B}^\prime$ is changed to match that of A. Since only the kinetic energy is changed in the second step, the free-energy difference can be evaluated analytically~\cite{Yang2018}:
\begin{equation}
    \Delta G = Nk_\textrm{B}T\left [ \frac{3}{2}x\ln\frac{m_\textrm{B}}{m_\textrm{A}}+x\ln x+(1-x)\ln(1-x)\right].
\end{equation}
This process only needs to be performed at one temperature, and Eq.~\ref{eq:gibbs-helmoltz} can be used to efficiently extend to other temperatures.

\begin{figure}[tb]
    \centering
    \includegraphics[width=\linewidth]{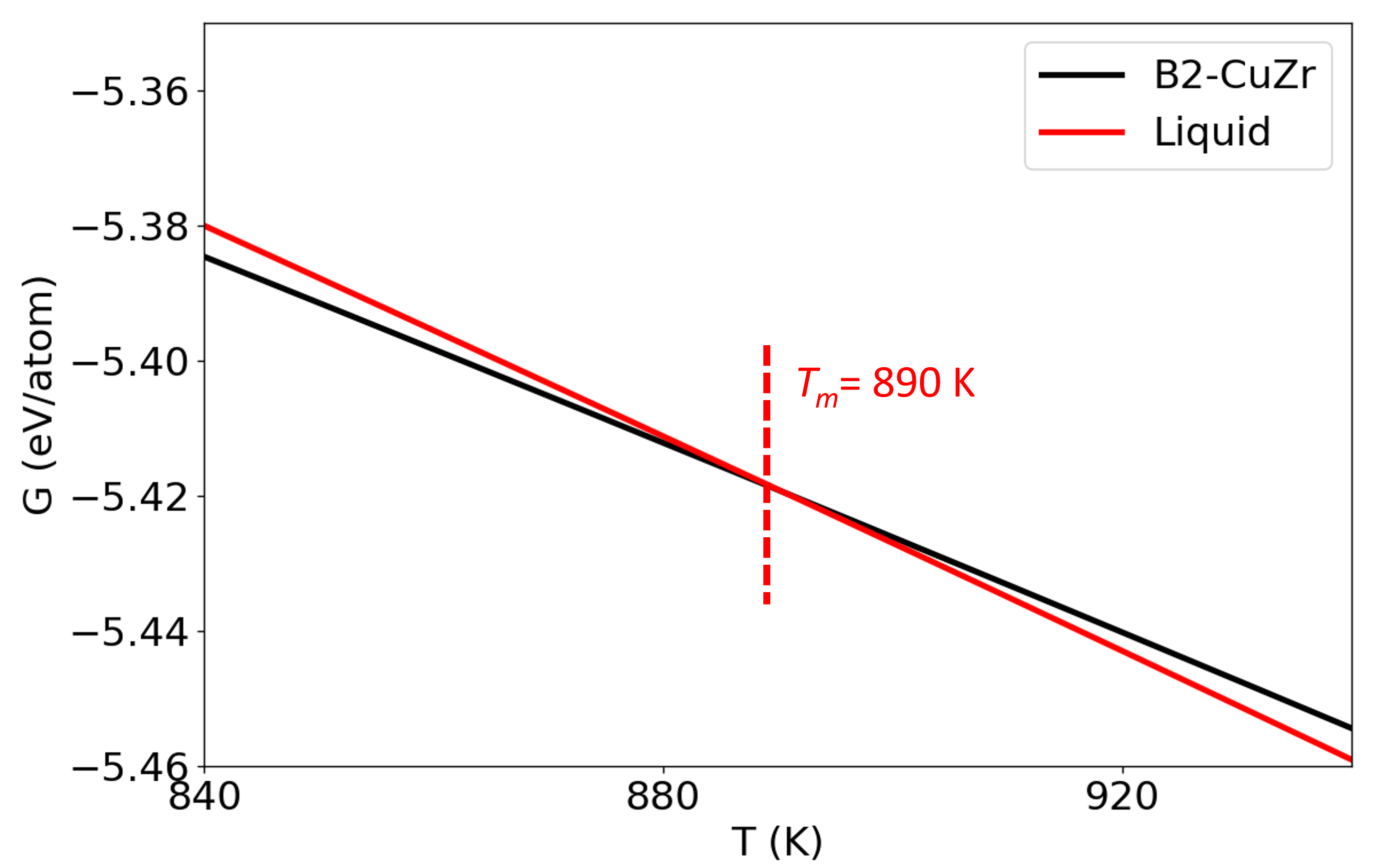}
    \caption{The Gibbs free energy of the CuZr B2 phase and the Cu$_{50}$Zr$_{50}$ liquid as a function of the temperature. The liquid free energy is calculated using thermodynamic integration along an alchemical pathway starting from the pure Cu liquid. The free energy of the B2 phase is calculated using the Einstein crystal as the reference. The crossing point gives the melting point $T_m=890$ K.}
    \label{fig:b2-cuzr}
\end{figure}

A practical way is to use UFM to obtain the free energy for pure A, and then use the alchemical TI to extend to other compositions $\textrm{A}_{1-x}\textrm{B}_x$. In Fig.~\ref{fig:b2-cuzr}, we show the $G$ vs. $T$ curve for the $\textrm{Cu}_{50}\textrm{Zr}_{50}$ liquid phase calculated using the alchemical TI approach. Also shown is the Gibbs free energy for the solid B2-CuZr phase derived from the Frankel-Ladd TI. The intersect of the two curves gives the melting temperature of $890$ K, which compares favorably with the value of 903 K measured independently using the Solid-liquid coexistence method, to be discussed in the next subsection (see Fig.~\ref{fig:sli-cuzr}). The deviation of 13 K, or 1.4\%, falls well within the expected accuracy of the current method.

\subsection{Solid-liquid coexistence}~\label{subsec:slc}
The workflow includes a module for measuring the melting temperature ($T_m$) using the solid-liquid coexistence (SLC) technique in which $T_m$ is determined by monitoring the migration of the solid-liquid interface~\cite{Morris1994}. As this method involves no underlying approximations, the SLC method is expected to yield very accurate $T_m$ values~\cite{Wilson2015}. On the other hand, it requires a large system, typically comprising $\sim$ 10,000 or more atoms, to model the solid-liquid interface effectively. Therefore, an efficient interatomic potential is required to implement methods, and it is in general not applicable for ab initio modeling. In theory, the SLC method can be used to simulate $T_m$ for both congruent and incongruent melting~\cite{Pedersen2018}, here we focus only on congruent melting in which the solid and liquid phases have the some composition, since it does not require long-range mass transport associated with the composition change in incongruent melting, and thus is more efficient. 


To implement the SLC method in the workflow, a supercell is generated with an aspect ratio of at least $2:1$. The system is first thermalized under the $NPT$ ensemble. Subsequently, one half of the supercell along the long axis (the default is $z$-axis) is melted by raising the temperature well above the melting point, while keeping the other half intact. Then, the liquid half is quickly quenched to the target temperature, and the solid half is released, resuming the integration of equations of motion for the whole system. Depending on the temperature, the interface will start moving toward the liquid side (solidification) or the solid side (melting). During the entire simulation except for the initial equilibration, an $NP_zT$ ensemble is implemented with a uniaxial barostat that only allows the dimension of the simulation box along the $z$-axis to change while the transverse dimensions remain fixed. The periodic boundary conditions are maintained throughout the simulation. The movement of the interface is monitored by tracking the total energy ($E$) of the system, which increases during melting and decreases during solidification due to the latent heat. This process is repeated at various temperatures, and interpolation of the measured rate of the total energy change can give $T_m$, where $dE/dt=0$.

\begin{figure}[tb]
    \centering
    \includegraphics[width=0.9\linewidth]{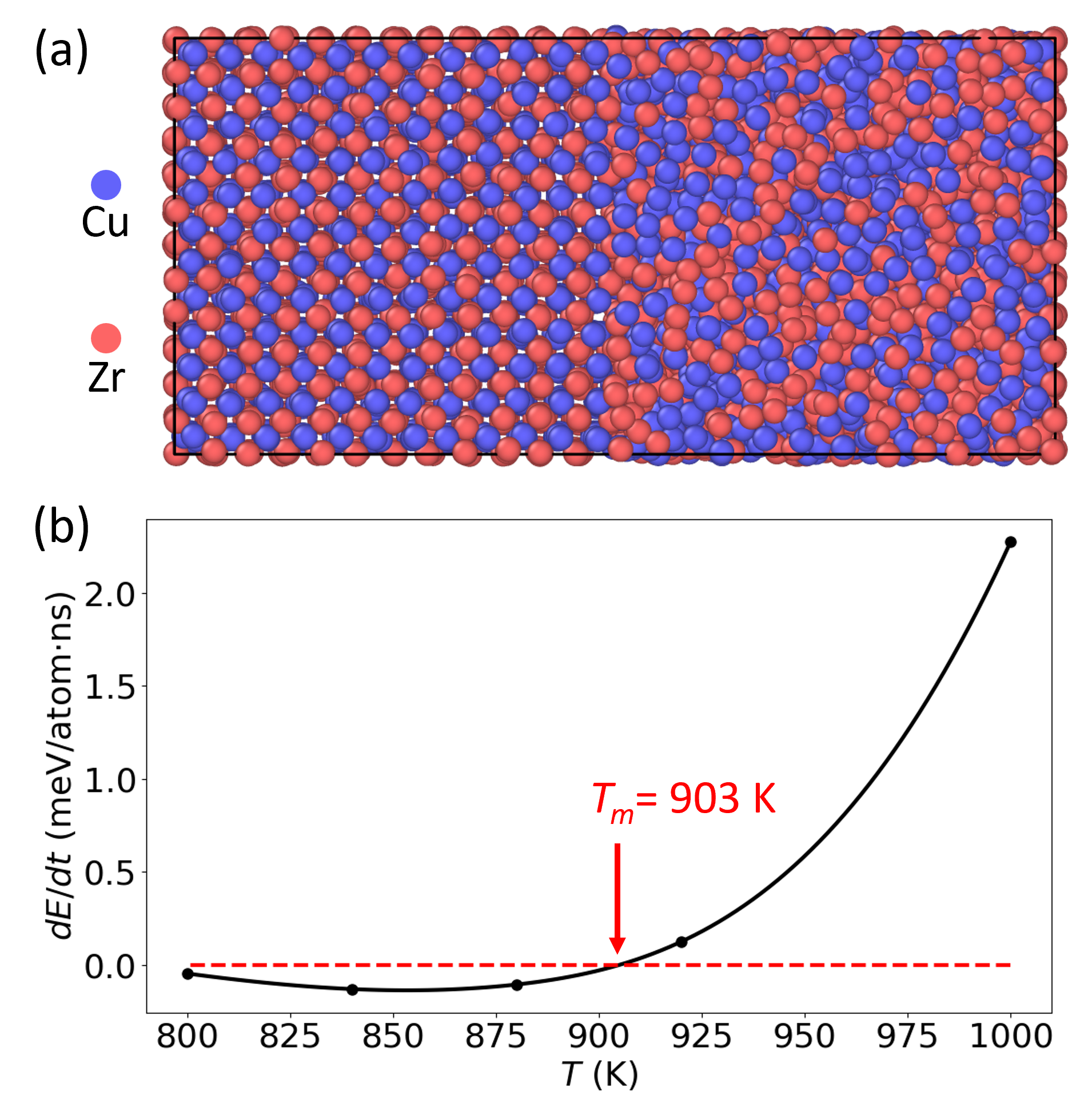}
    \caption{Measurement of the melting point of the Cu-Zr B$_2$ phase using the SLC method. (a) The initial configuration of the solid-liquid interface at $T=800$ K. On the left is the crystalline Cu-Zr B$_2$ phase and on the left is the liquid structure with $x_\textbf{Zr}=0.5$. (b) The rate of the internal energy change as a function of the temperature during the melting or the crystallization process.  The solid line is a cubic interpolation, which gives the melting point $T_m=903$ K when the interpolated $\frac{\partial E}{\partial t}=0$.}
    \label{fig:sli-cuzr}
\end{figure}

We demonstrate the method's broader applicability by selecting a binary B2-CuZr phase. Fig.~\ref{fig:sli-cuzr} (a) illustrates the configuration of the solid-liquid interface after the liquid phase has been quenched to a target temperature of $T=800$ K. The left side shows the B2 structure, while the right side is the amorphous liquid with the same composition of $x_\textrm{Zr}=0.5$. In Fig.~\ref{fig:sli-cuzr} (b), the solid circles give the rate of the total energy change measured in MD simulations across various temperatures, and the solid line is a cubic interpolation. The intersection with the dashed red line gives $T_m=903$ K. It can be noted that below $T_m$, $|dE/dt|$ indicative of the crystal growth rate reaches a maximum at $T\sim860$ K. Above this temperature, the driving force for crystallization, defined as the free energy difference between solid and liquid phases, is small; while below this temperature, the kinetics becomes sluggish.

Although the current workflow does not use the SLC method for free energy calculations, SLC simulations are helpful to validate the free energy results. For example, the agreement between the melting point of the B2-CuZr phase derived from SLC simulations (Fig.~\ref{fig:sli-cuzr}
) and from free energy calculations (Fig.~\ref{fig:b2-cuzr} confirms the reliability of both methods. Furthermore, SLC simulations can be used to study crystal growth kinetics, which will be a primary focus in Phase II of the workflow development.

\subsection {CALPHAD}
\begin{figure}[tb]
    \centering
    \includegraphics[width=0.9\linewidth]{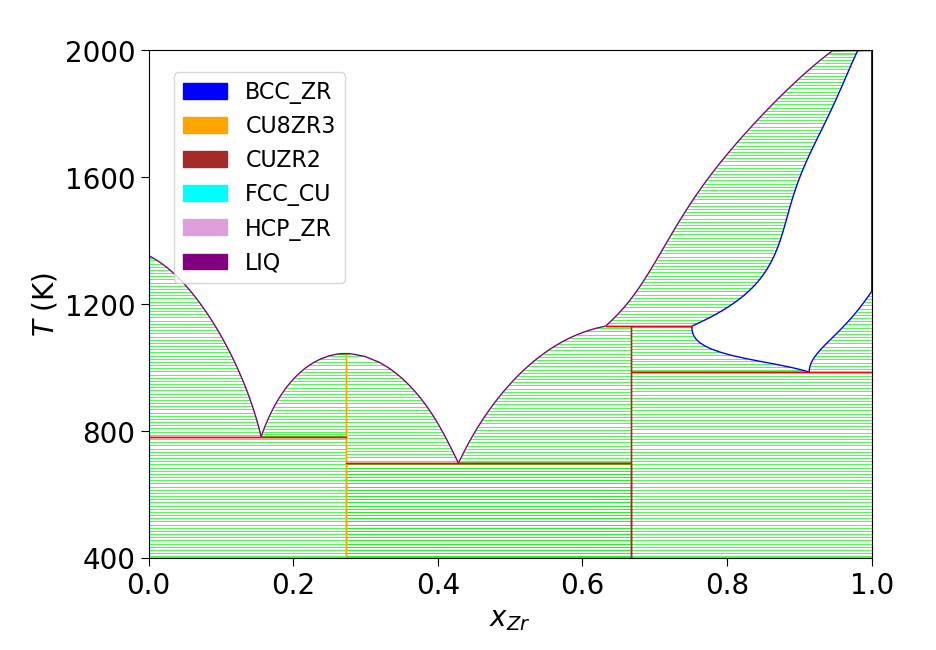}
    \caption{Phase diagram of the Cu-Zr system at ambient pressure calculated using an EAM-FS potential~\cite{Mendelev2019}. The phases involved in the phase-diagram calculation include fcc-Cu, hcp-Zr, bcc-Zr, $\textrm{Cu}_5\textrm{Zr}$, $\textrm{Cu}_8\textrm{Zr}_3$, $\textrm{Cu}_{10}\textrm{Zr}_7$, and $\textrm{Cu}\textrm{Zr}_2$, and the liquid phase. The solubility of Cu in the hcp-Zr phase is not considered. }
    \label{fig:pd-cuzr}
\end{figure}

As mentioned earlier, the Gibbs free energy of a non-ideal solution is represented by the Redlich-Kistler polynomial given in Eqs.~\ref{eq:sol} and~\ref{eq:rk}. In addition, all functions of temperatures, including the Gibbs free energy of line compounds as we discussed in Subsection~\ref{subsec:linecompount}, the Gibbs free energy of the end members $G_0$ and $G_1$ in Eq.~\ref{eq:sol}, and the coefficients of the Redlich-Kistler polynomial $L_i$ (Eq.~\ref{eq:rk}), are often fitted to the following form:
\begin{equation}\label{eq:polylog}
    G(T)=cT\ln{T}+\sum_{n=-1}^{n_\textrm{max}}d_nT^n.
\end{equation}
In practice, we set $n_\textrm{max}=3$, resulting in a total of 6 fitting parameters in Eq.~\ref{eq:polylog}. In this way, both solution and non-solution phases can be represented with a handful of parameters, which are then grouped into a thermodynamic database in the standard TDB format developed by Thermo-Calc~\cite{Andersson2002}. To identify phases in equilibrium is equivalent to minimizing the total Gibbs free energy $G_\textrm{tot}=\sum\limits_{\phi=1}^{N_\phi} G^\phi$, subject to the constraints that the total content of each element conserves: $\sum\limits_{\phi=1}^{N_\phi} n^\phi x_i^\phi=n_i$ for any $1\leq i\leq N_{el}$; and the net composition of each phase is one: $\sum\limits_{i=1}^{N_{el}}x_i^\phi=1$ for any $1\leq\phi\leq N_\phi$. Here, $\phi$ indexes the phases and $i$ indexes the elements. $n^\phi$ is the number of moles of phase $\phi$, $x_i^\phi$ is the mole fraction of element $i$ in phase $\phi$, and $n_i$ is the total number of moles of element $i$ in the mixture. In our workflow, we create the thermodynamic database file from the free energy calculations and then implement the open-source package PYCalphad~\cite{Otis2017} to solve the optimization problem with constraints and obtain the phase diagram. In Fig.~\ref{fig:pd-cuzr}, we show the phase diagram of the Cu-Zr system at zero pressure, calculated using a newly developed EAM-FS potentail~\cite{Mendelev2019}. The solid phases fcc-Cu, hcp-Zr, bcc-Zr, $\textrm{Cu}_5\textrm{Zr}$, $\textrm{Cu}_8\textrm{Zr}_3$, $\textrm{Cu}_{10}\textrm{Zr}_7$, and $\textrm{Cu}\textrm{Zr}_2$ are included in the phase-diagram calculation, together with the liquid phase. Compared to an earlier EAM-FS potential~\cite{Mendelev2009}, the new potential corrects the unphysical stability of the B2-CuZr phase and a Laves Cu$_2$Zr phase~\cite{Tang2012}. While there remains no consensus on the experimental Cu-Zr phase diagram since different thermodynamic assessments often yield different results~\cite{Okamoto2012}, the new potential still exhibits two notable deficiencies: it creates a too deep eutectic points in the Cu-rich region; and it overestimates the solubility of Cu in the bcc-Zr phase (the solubility of Cu in the hcp-Zr phase is not considered in the current calculations)~\cite{Okamoto2012}. 

\section{Scalable task-based parallel workflow execution with Parsl}
exaPD leverages Parsl, a parallel programming library for Python, to scale the workload of hundreds of MD jobs with internal dependencies across heterogeneous resources on large-scale computational systems. By abstracting task execution into a flexible dependency graph, Parsl enables a data-driven execution model in which tasks are triggered as soon as their inputs become available. While most MD jobs leverage GPU acceleration, certain essential features are CPU-only, necessitating a heterogeneous CPU/GPU architecture. A Parsl executor is configured for each resource type and tasks are assigned to the corresponding executors based on their type. Parsl allows researchers to build modular, task-based execution pipelines that scale seamlessly from local machines to high-performance computing clusters. In this work, calculations were performed on a large-scale cluster system using Slurm; however, Parsl also supports cloud platforms and other cluster management systems. Transitioning between different environments requires only minor adjustments to the Parsl configuration file.

\begin{figure}[tb]
    \centering
    \includegraphics[width=\linewidth]{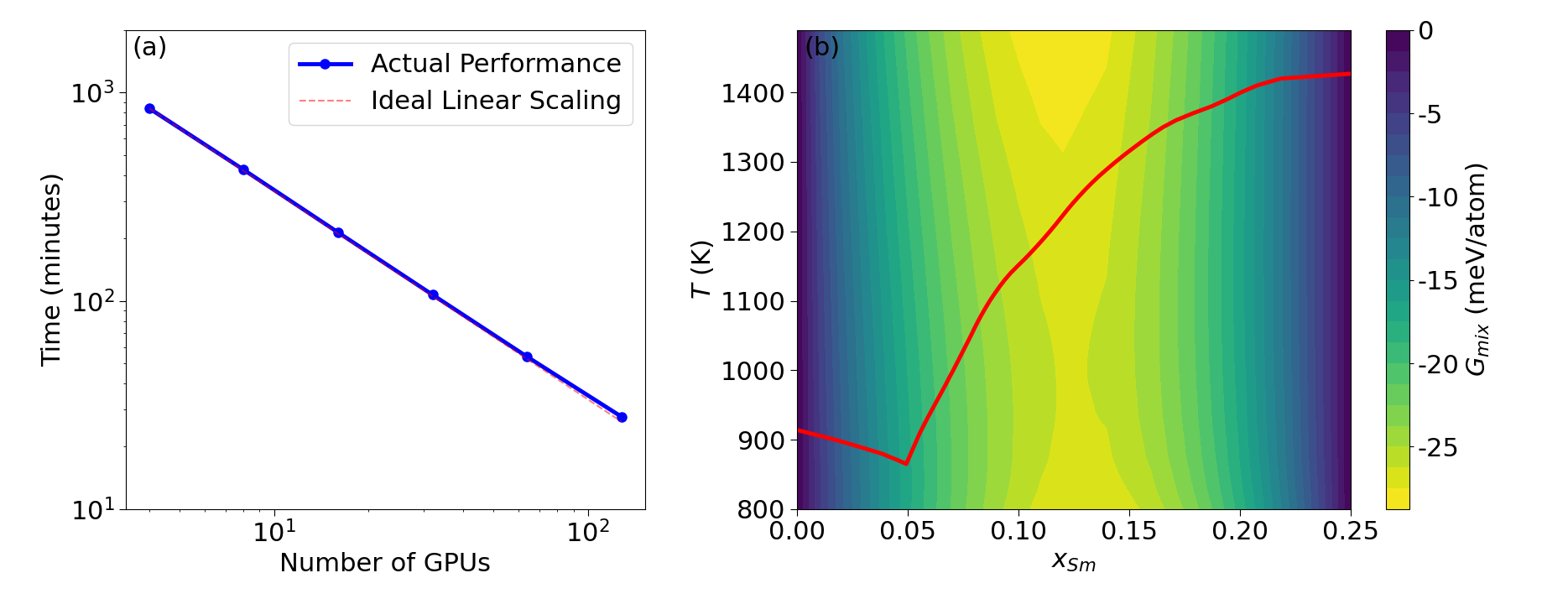}
    \caption{(a) Running time as a function of the number of GPUs for the task of calculating the free energy of Al-Sm liquid ($x_\textbf{Sm}\leq0.25$) for a wide range of temperatures. The dashed line shows the ideal strong scaling. (b) Contour plot of the Al-Sm liquid free energy referenced to pure Al liquid and Al$_{0.75}$Sm$_{0.25}$ liquid. The red curve is the melting line showing a eutectic point between two solid phases fcc-Al and Al$_3$Sm. }
    \label{fig:scale}
\end{figure}

As an example, we demonstrate in Fig.~\ref{fig:scale} the scalability of the workflow in the free energy calculating of the Al-Sm liquid in the Al-rich regime ($0\leq x_\textrm{Sm}\leq0.25$), using an EAM-FS potential~\cite{Mendelev2015}. Hundreds of MD jobs are required to map out the Gibbs free energy as a function of $T$ and $x_\textrm{Sm}$. Fig.~\ref{fig:scale} plots the total run time as a function of the number of GPUs used for the calculation, which shows almost ideal strong scaling. The results are presented in Fig.~\ref{fig:scale} (b). Here, the mixing free energy $G_{mix}$ is defined using the free energy at two limiting compositions $x_\textrm{Sm}=0$ and $x_\textrm{Sm}=0.25$ as reference: $G_{mix}(x,T)=G(x,T)-[(1-4x)G(0,T)+4xG(0.25,T)$. The free energy of the liquid phase, combined with the free energy for two solid phases fcc-Al and $\textrm{Al}_3\textrm{Sm}$, which was calculated separately, produces the melting curve (red) for the two solid phases. 

\section{Structure of the workflow and the user interface}
Below, we list the major modules for laying out all necessary MD jobs for the construction of the phase diagram and 2 modules to define a Parsl configuration for running workflows with both GPU and CPU resources.
\begin{itemize}
    \item \path{einstein.py}: It sets up the Frankel-Ladd TI for solids using an Einstein crystal as the reference system. The $NVT$ ensemble is used in this procedure.  It requires a prerequisite process to equilibrate the system at the target temperature and pressure to obtain the equilibrium box size as well as the MSD for each element; the latter will be used to determine the spring constants for the Einstein crystal. 
    
    \item \path{alchem.py}: This module is used in the workflow to set up TI calculations to transform a pure liquid to a target liquid alloy. If the user wants the alloy AB to interact in the same way as it does in the pure system A, the user should specify in the input JSON file (will be discussed below) how it is achieved in LAMMPS script. For example, for an LJ system, this is done via ``\texttt{pair\_coeff * * $\epsilon_\textrm{AA}$ $\sigma_\textrm{AA}$}", while for an EAM-FS potential, one can write: ``\texttt{pair\_coeff * * AB.eam.fs A A}". If the script for defining how the pure system interacts is not provided explicitly, the UFM will be used instead. The $NVT$ ensemble will be used if the default UFM is used as reference. For this reason, a pre-equilibration procedure is required to obtain the equilibrium volume. Otherwise, the $NPT$ ensemble will be used instead. In addition to its function in the current workflow for phase diagram calculations, it can also serve the general purpose of transforming one type of interatomic potential to another type for either the solid or liquid phase. As an example, we have demonstrated how to calculate the Gibbs free energy of fcc-Al with the NNP from the EAM-FS potential in Fig.~\ref{fig:fcc-al}.
    
    \item \path{tramp.py}: This module is used to ramp up or ramp down the temperature for solid or liquid phases. The $NPT$ ensemble is used according to the target temperature and pressure. The enthalpy $H$ as a function of $T$ is obtained in this procedure, which is used in the Gibbs-Helmholtz integration in Eq.~\ref{eq:gibbs-helmoltz} to extend the Gibbs free energy calculated at one temperature using TI to other temperatures. At certain temperatures, this step also provides the prerequisite parameters for other processes as described above. 

    \item \path{sli.py}: This is an optional module for determining the melting point of a certain solid phase using the SLC method. If this process is included, then no other reference system is required for the liquid phase. Otherwise, the UFM will be used to obtain the absolute free energy for the liquid. During equilibration, only the dimension perpendicular to the interface is allowed to change, while the simulation box along the transverse directions is fixed according to a pre-equilibration MD job.

    \item \path{sgmc.py}: This is also an optional module that uses the semi-grand canonical Monte Carlo method to calculate the Gibbs free energy of a solid solution phase. This is only required if there is a relevant solid solution phase in the system. Extra cautions are required for setting a proper range of the chemical potential difference ($\Delta\mu$) at each temperature [see Fig.~\ref{fig:sgmc-bcc} (a)]. If $\Delta\mu$ is too small, the solid will saturate on one end; on the other hand, if $\Delta\mu$ is too large, it results in an unrealistically large alloying level that causes the solid phase to melt. Considering the usually strong non-linear nature of the $\Delta\mu$ vs. $x$ curve, a non-equidistant list of $\Delta\mu$ is preferred.

    \item \path{config_loader.py}: This module loads a Parsl configuration object based on a user-specified name in the runtime configuration dictionary, which defines two executors for running workflows: one for GPU jobs and one for CPU jobs.

    \item \path{lammps.py}: This module defines two Parsl bash\_app functions to run LAMMPS jobs either on GPU or CPU resources. Both apps take as input the working directory, the LAMMPS input script, and the executable path, and they generate a shell command string that Parsl executes.
\end{itemize}
The Nose-Hoover thermostat and/or barostat is used in all the above modules, except for \path{einstein.py}, in which the Langevin thermostat is used due to the instability of the Nose-Hoover thermostat in treating Harmonic degrees of freedom.

All the input data is arranged in a JSON file, which is made up of five parts, ``\texttt{general}", ``\texttt{run}", ``\texttt{liquid}", ``\texttt{solid}", ``\texttt{sli}", ``\texttt{sgmc}", with the last two being optional. Below we describe the function as well as the required and functional settings in each component. 

\begin{itemize}
    \item ``\texttt{general}" determines the target system and the global settings for the LAMMPS calculation.
    \begin{itemize}
        \item Required settings
        \begin{itemize}
            \item ``\texttt{system}": a string of all the elements of the system separated by space.
            \item ``\texttt{mass}": a list of masses for each element.
            \item ``\texttt{pair\_style}": the pair\_style in LAMMPS syntax that defines the interatomic potential.
            \item ``\texttt{pair\_coeff}": the pair\_coeff associated with the pair\_style in LAMMPS syntax. It can be a single line or a list of multiple lines.
        \end{itemize}
        \item Optional settings
        \begin{itemize}
            \item ``\texttt{proj\_dir}": the path to the root directory of the project for running the calculations. Default is the current directory.
            \item ``\texttt{pressure}": the target pressure. Default is 0.
            \item ``\texttt{units}": the units for LAMMPS calculation, ``\texttt{metal}" or ``\texttt{lj}" are supported. Default is ``\texttt{metal}".
            \item ``\texttt{timestep}": the timestep for MD calculations. Default is 0.001 for ``\texttt{metal}" units and 0.005 for ``\texttt{lj}" units.
            \item ``\texttt{run}": the total number of steps to run in MD calculations. Default is $10^6$.
            \item ``\texttt{Tdamp}": the time period for temperature damping in thermostating. Default is 100$\times$``\texttt{timestep}".
            \item ``\texttt{Pdamp}": the time period for pressure damping in barostating. Default is 1000$\times$``\texttt{timestep}".
            \item ``\texttt{thermo}": the number of timesteps between two consecutive outputs in MD simulations. Default is 100.
        \end{itemize}
    \end{itemize}
\end{itemize}

\begin{itemize}
    \item ``\texttt{run\_config}" provides run-time parameters for launching LAMMPS jobs using Parsl. It includes both GPU and CPU execution options, as well as scheduler directives. The configuration provided in this example targets systems that use Slurm as the workload manager. For environments with different schedulers or non-scheduler setups (e.g., local machines, cloud platforms), users may customize the Parsl configuration by replacing the SlurmProvider with the appropriate provider or executor settings.
    \begin{itemize}
        \item Required settings
        \begin{itemize}
            \item ``\texttt{ngpu}": the number of nodes required for each GPU job submitted by Parsl. Default is 1.
            \item ``\texttt{ncpu}": the number of nodes required for each CPU job submitted by Parsl. Default is 1.    
            \item ``\texttt{gpu\_exe}": the executable command or path to run LAMMPS on GPU resources.
            \item ``\texttt{cpu\_exe}": the executable command or path to run LAMMPS on CPU resources.
            \item ``\texttt{parsl\_config}": the Parsl configuration profile that specifies how jobs are launched and resources are allocated.
        \end{itemize}
        \item Optional settings
        \begin{itemize}
            \item ``\texttt{gpu\_schedule\_option}": a list of Slurm scheduler directives used when launching GPU jobs. These options define constraints such as GPU architecture, walltime, account, GPU allocation per node, and queue. Default is null.
            \item ``\texttt{cpu\_schedule\_option}": a list of Slurm scheduler directives used when launching CPU jobs. Similar to the GPU case, but targeting CPU-only nodes. Default is null.
        \end{itemize}
    \end{itemize}
\end{itemize}

\begin{itemize}
    \item ``\texttt{liquid}" determines extra parameters for liquid free energy calculations.
    \begin{itemize}
        \item Required settings
        \begin{itemize}
        \item ``\texttt{data\_in}": input data file for the liquid structure in the atom style of the LAMMPS data format. Ensure that no atoms are unphysically close to one another. The atom types are not important, as they will be modified during the alchemical process.
            \item ``\texttt{initial\_comp}": the initial composition for the alchemical process.
            \item ``\texttt{final\_comp}": the final composition for the alchemical process.
            \item (``\texttt{Tmin}", ``\texttt{Tmax}" and ``\texttt{dT}") and ``\texttt{Tlist}": the former refers to the minimal temperature, the maximal temperature, and the temperature increment, while the latter is a list of temperatures. At least one of these two sets of parameters needs to be provided. If both are provided, a sorted temperature list will be generated by combining them and removing duplicates. This feature is helpful for setting non-equidistant temperatures or for adding additional temperatures after the initial run. 
        \end{itemize}
        \item Optional settings
        \begin{itemize}
            \item ``\texttt{ncomp}": the number of compositions in between the initial and final compositions. Default is 10. 
            \item ``\texttt{ref\_pair\_style}" and ``\texttt{ref\_pair\_coeff}": The pair style and coefficient defining the reference system. Default is the UFM.
            \item ``\texttt{dlbd}": $\Delta\lambda$ used in TI. Default is 0.05.
        \end{itemize}
    \end{itemize}
\end{itemize}

\begin{itemize}
    \item ``\texttt{solid}" determines extra parameters for liquid free energy calculations.
    \begin{itemize}
        \item Required settings
        \begin{itemize}
            \item  ``\texttt{phases}": list of solid phases (line compounds) for free energy calculations. It accepts unit-cell structures in popular formats such as CIF or VASP. It also accepts the standard lammps input file with the extension ``\texttt{.lammps}". If a unit-cell structure is provided, the ASE package~\cite{Sander2017} will be used to generate a supercell containing $\sim$ 5000 atoms. Also, if the structure if triclinic or monoclinic, it is the user's responsibility to create a ``cubic"-like box for MD runs.
            \item (``\texttt{Tmin}", ``\texttt{Tmax}" and ``\texttt{dT}") or ``\texttt{Tlist}": the same as in ``\texttt{liquid}".
        
        \end{itemize}
        \item Optional settings
        \begin{itemize}
            \item ``\texttt{dlbd}": the same as in ``\texttt{liquid}".
            \item ``\texttt{ntarget}": the target size of the supercell for solid structures. The program will generate a supercell with the number of atoms close to ``\texttt{ntarget}" for each solid phase.
        \end{itemize}
    \end{itemize}
\end{itemize}

\begin{itemize}
    \item ``\texttt{sli}" determines extra parameters for solid-liquid interface (SLI) simulations.
    \begin{itemize}
        \item Required settings
        \begin{itemize}
            \item  ``\texttt{phases}": the same as in ``\texttt{solid}.
            \item (``\texttt{Tmin}", ``\texttt{Tmax}" and ``\texttt{dT}") or ``\texttt{Tlist}": the same as in ``\texttt{liquid}".
            \item ``\texttt{Tmelt}": a high temperature to melt half of the solid phase to prepare a SLI.
        \end{itemize}
        \item Optional settings
        \begin{itemize}
            \item ``\texttt{orientation}": the orientation of the SLI, which takes the value of ``\texttt{x}", ``\texttt{y}" or ``\texttt{z}". The default value is ``\texttt{z}".
            \item ``\texttt{ntarget}": the same as in ``\texttt{solid}".
            \item ``\texttt{replicate}": the number of replicates of the supercell along the ``\texttt{orientation}" direction, half of which is melted at the beginning of the simulation to create a SLI. The default value is 2. 
        \end{itemize}
    \end{itemize}
\end{itemize}

\begin{itemize}
    \item ``\texttt{sgmc}" determines extra parameters for semi-grand canonical ensemble calculations.
    \begin{itemize}
        \item Required settings
        \begin{itemize}
            \item  ``\texttt{phases}": the same as in ``\texttt{solid}.
            \item (``\texttt{Tmin}", ``\texttt{Tmax}" and ``\texttt{dT}") or ``\texttt{Tlist}": the same as in ``\texttt{liquid}".
            \item (``\texttt{mu\_min}", ``\texttt{mu\_max}" and ``\texttt{dmu}") or ``\texttt{mu\_list}": determines a list of $\mu\equiv\mu_A-\mu_B$. It behaves in the same way as temperature settings described in ``\texttt{liquid}".
        \end{itemize}
    \end{itemize}
\end{itemize}

\begin{figure}[tb]
    \centering
    \includegraphics[width=0.9\linewidth]{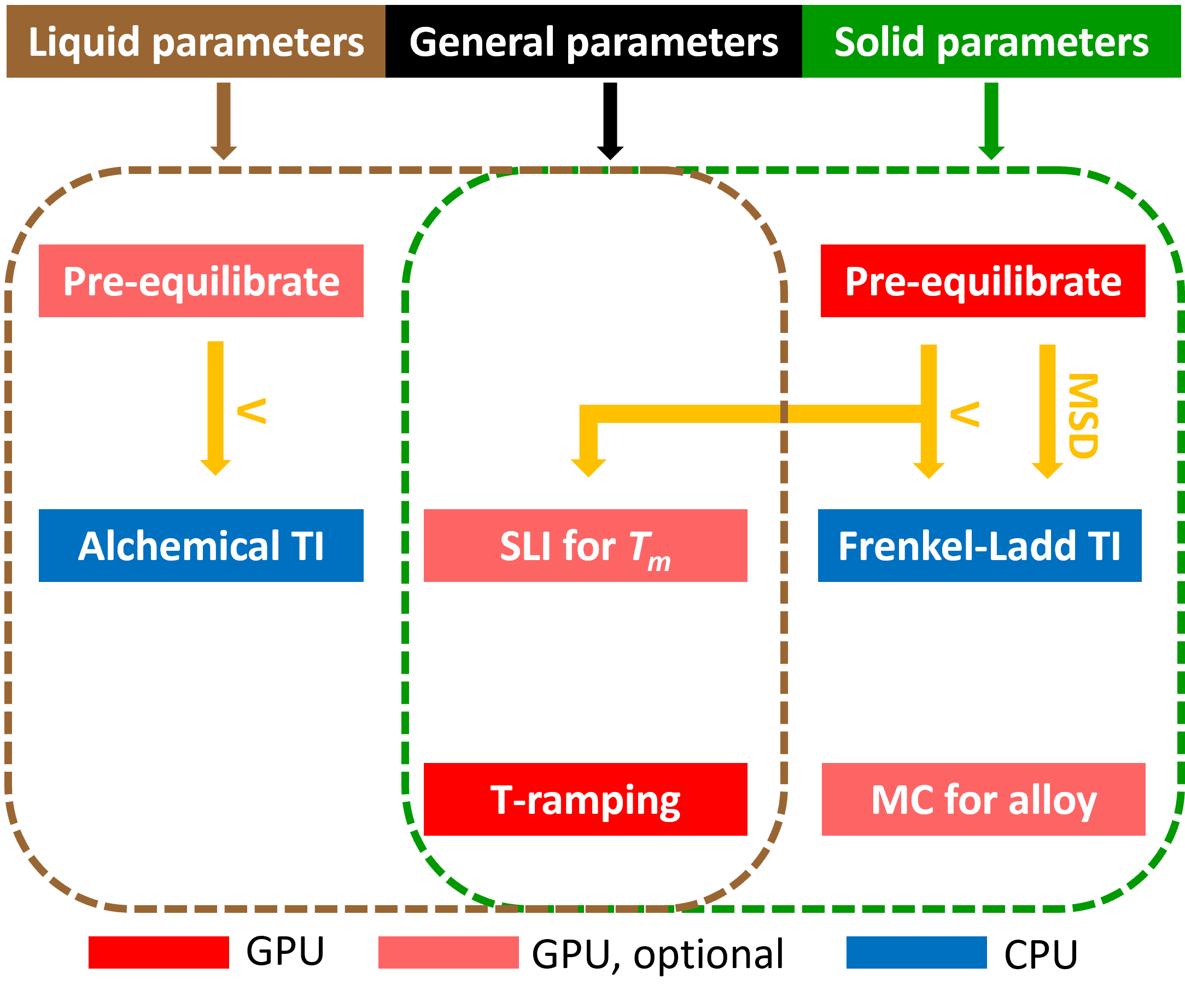}
    \caption{Schematic flowchart of the exaPD workflow.}
    \label{fig:flowchart}
\end{figure}

Fig.~\ref{fig:flowchart} gives a schematic flowchart of the exaPD workflow, outlining the required and optional MD jobs for constructing a phase diagram, taking inputs that are general to the entire project or are specific to the solid or liquid simulations. It also illustrates the internal dependencies among the MD jobs. For instance, the solid phase needs to be pre-equilibrated to obtain the equilibrium volume and the MSD for each speicies, in order to set up the Frenkel-Ladd TI with the einstein crystal as the reference state. The dependences are managed by the Parsl controller using futures. 

By default, the Frenkel-Ladd and alchemical TI calculations are performed on CPUs, as certain LAMMPS features for these calculations are not currently supported by GPU or KOKKOS, the two primary packages for GPU acceleration. However, when using the pre-compiled LAMMPS executable from in the DeepMD package to implement the DeepMD NNP, all the calculations shown in Fig.~\ref{fig:flowchart} can be GPU-accelerated. In this case, users can override the default setting by assigning the value ``\texttt{gpu}" to the ``\texttt{\_arch}" feature of all the jobs in the main program \path{run.py}, which is executed to send all the MD jobs to the job scheduler of the computing system. 

Each job runs in a separate directory, and an empty file \path{DONE} will be generated in the directory after the job is completed normally. If all the jobs are not finished in the initial run due to the wall-time limit, or if new jobs are added (e.g., to expand the temperature range), one can edit the configuration JSon file accordingly and rerun \path{run.py}. Only unfinished or new jobs will be submitted. To perform fresh calculations for specific jobs, users must clear the corresponding directories before re-executing \path{run.py}.

Finally, \path{run_process.py} is the program to post-process the calculations. It generates a two-column data file of $G$ versus $T$ for each solid phase, and a multi-column data file of $G$ versus $T$ and $x$ for the liquid phase, with each composition in a separate column. In addition, it creates a thermodynamic database file in the TDB format, containing entries for the calculated solid and liquid phases. A sample \path{plot_PD.py} script is provided to plot the phase diagram for the Cu-Zr system using the PyCALPHAD package based on the TDB file. For advanced thermodynamic calculations using the database, users are referred to the PyCALPHAD documentation~\cite{pycalphad2025}.

In general, post-processing the semi-grand canonical ensemble and the solid-liquid coexistence simulations involves monitoring the simulation process using visualization tools. Thus, a generic post-processing script is not currently provided for these two optional modules \path{sli.py} and \path{sgmc.py}. Users can refer to Subsections~\ref{subsec:sgmc} and~\ref{subsec:slc} for analyzing these simulations.

\section{Code availability}

The code is publicly available at https://github.com/ML-AMD/exa-pd.

\section{Conclusion}
We present exaPD, a user-friendly package for the computational study of phase diagrams. It provides a highly scalable workflow for accurate free energy calculations across a wide range of temperatures and compositions. By integrating standard sampling techniques such as molecular dynamics (MD) and Monte Carlo (MC) through the LAMMPS package, exaPD supports various interatomic potentials, including highly accurate neural network potentials, enabling precise simulations of complex materials. The implementation of a global controller using Parsl ensures massive parallelization with near-ideal scalability, efficiently managing MD/MC jobs to handle resource-intensive calculations. Coupled with CALPHAD modeling, exaPD facilitates the generation of reliable phase diagrams. Future development phases will incorporate nucleation and growth kinetics, as well as liquid structure analysis, which are key factors in phase selection during liquid-based synthesis. The ultimate goal is to establish a robust framework that empowers researchers to acquire thermodynamic and kinetic data in a timely manner on exascale computing facilities, guiding the synthesis of advanced materials with enhanced accuracy and efficiency.

\section*{Acknowledgements}
Work at Ames National Laboratory and Los Alamos National Laboratory was supported by the U.S. Department of Energy (DOE), Office of Science, Basic Energy Sciences, Materials Science and Engineering Division through the Computational Material Science Center program. Ames National Laboratory is operated for the U.S. DOE by Iowa State University under contract No. DE-AC02-07CH11358. Los Alamos National Laboratory is operated by Triad National Security, LLC, for the National Nuclear Security Administration of U.S. Department of Energy under Contract No. 89233218CNA000001. This
research used resources of the National Energy Research Scientific Computing Center (NERSC), a DOE Office of Science User Facility supported under Contract No. DE-AC02-05CH11231. (LA-UR-25-28627)

\bibliographystyle{elsarticle-num}

\begin{thebibliography}{10}
\expandafter\ifx\csname url\endcsname\relax
  \def\url#1{\texttt{#1}}\fi
\expandafter\ifx\csname urlprefix\endcsname\relax\def\urlprefix{URL }\fi
\expandafter\ifx\csname href\endcsname\relax
  \def\href#1#2{#2} \def\path#1{#1}\fi

\bibitem{Gubernatis2021}
J.~E. Gubernatis, T.~Lookman, Machine learning in materials design and discovery: examples from the present and suggestions for the future, Physical Review Materials 129~(7) (2021) 070401.

\bibitem{Vasudevan2021}
R.~Vasudevan, G.~Pilania, P.~V. Balachandran, Machine learning for materials design and discovery, Journal of Applied Physics 129~(7) (2021) 070401.

\bibitem{Kusne2014}
A.~G. Kusne, T.~Gao, A.~Mehta, L.~Q. Ke, M.~C. Nguyen, K.~M. Ho, V.~Antropov, C.~Z. Wang, M.~J. Kramer, C.~Long, I.~Takeuchi, On-the-fly machine-learning for high-throughput experiments: search for rare-earth-free permanent magnets, Scientific Reports 4 (2014) 6367.

\bibitem{Kabiraj2020}
A.~Kabiraj, M.~Kumar, S.~Mahapatra, High-throughput discovery of high curie point two-dimensional ferromagnetic materials, npj Computational Materials 6 (2020) 35.

\bibitem{cai2020}
J.~Cai, X.~Chu, K.~Xu, H.~Li, J.~Wei, Machine learning-driven new material discovery, Nanoscale Advances 2 (2020) 3115.

\bibitem{Katsikas2021}
G.~Katsikas, S.~Charalampos, K.~Joseph, Machine learning in magnetic materials, Physica Status Solidi (b) 258 (2021) 2000600.

\bibitem{Rhone2020}
T.~D. Rhone, W.~Chen, S.~Desai, S.~B. Torrisi, D.~T. Larson, A.~Yacoby, E.~Kaxiras, Data-driven studies of magnetic two-dimensional materials, Scientific Reports 10 (2020) 15795.

\bibitem{Landrum2003}
G.~A. Landrum, H.~Genin, Application of machine-learning methods to solid-state chemistry: Ferromagnetism in transition metal alloys, Journal of Solid State Chemistry 176 (2003) 587.

\bibitem{Merchant2023}
A.~Merchant, S.~Batzner, S.~S. Schoenholz, et~al., Scaling deep learning for materials discovery, Nature 624 (2023) 80--85.

\bibitem{Szymanski2023}
N.~J. Szymanski, et~al., An autonomous laboratory for the accelerated synthesis of novel materials, Nature 624 (2023) 86--91.

\bibitem{Mroz2022}
A.~M. Mroz, et~al., Into the unknown: How computation can help explore uncharted material space, Journal of the American Chemical Society 144~(41) (2022) 18730--18743.

\bibitem{Aykol2019}
M.~Aykol, V.~I. Hegde, L.~Hung, et~al., \href{https://doi.org/10.1038/s41467-019-10030-5}{Network analysis of synthesizable materials discovery}, Nature Communications 10 (2019) 2018.
\newblock \href {https://doi.org/10.1038/s41467-019-10030-5} {\path{doi:10.1038/s41467-019-10030-5}}.
\newline\urlprefix\url{https://doi.org/10.1038/s41467-019-10030-5}

\bibitem{Chen2024}
C.~Chen, D.~T. Nguyen, S.~J. Lee, N.~A. Baker, A.~S. Karakoti, L.~Lauw, C.~Owen, K.~T. Mueller, B.~A. Bilodeau, V.~Murugesan, M.~Troyer, \href{https://arxiv.org/abs/2401.04070}{Accelerating computational materials discovery with artificial intelligence and cloud high-performance computing: from large-scale screening to experimental validation} (2024).
\newblock \href {http://arxiv.org/abs/2401.04070} {\path{arXiv:2401.04070}}.
\newline\urlprefix\url{https://arxiv.org/abs/2401.04070}

\bibitem{Behler2017}
J.~Behler, \href{https://doi.org/10.1002/anie.201703114}{First principles neural network potentials for reactive simulations of large molecular and condensed systems}, Angewandte Chemie International Edition 56~(42) (2017) 12828--12840.
\newblock \href {https://doi.org/10.1002/anie.201703114} {\path{doi:10.1002/anie.201703114}}.
\newline\urlprefix\url{https://doi.org/10.1002/anie.201703114}

\bibitem{Blank2019}
T.~B. Blank, S.~D. Brown, A.~W. Calhoun, D.~J. Doren, \href{https://doi.org/10.1038/s41467-019-10343-5}{Physically informed artificial neural networks for atomistic modeling of materials}, Nature Communications 10 (2019) 2339.
\newblock \href {https://doi.org/10.1038/s41467-019-10343-5} {\path{doi:10.1038/s41467-019-10343-5}}.
\newline\urlprefix\url{https://doi.org/10.1038/s41467-019-10343-5}

\bibitem{Zhang2018}
L.~Zhang, J.~Han, H.~Wang, R.~Car, W.~E, \href{https://doi.org/10.1103/PhysRevLett.120.143001}{Deep potential molecular dynamics: A scalable model with the accuracy of quantum mechanics}, Physical Review Letters 120~(14) (2018) 143001.
\newblock \href {https://doi.org/10.1103/PhysRevLett.120.143001} {\path{doi:10.1103/PhysRevLett.120.143001}}.
\newline\urlprefix\url{https://doi.org/10.1103/PhysRevLett.120.143001}

\bibitem{Kirkwood1935}
J.~G. Kirkwood, Statistical mechanics of fluid mixtures, The Journal of Chemical Physics 3 (1935).
\newblock \href {https://doi.org/10.1063/1.1749657} {\path{doi:10.1063/1.1749657}}.

\bibitem{Frenkel2023}
D.~Frenkel, B.~Smit, Understanding Molecular Simulation: From Algorithms to Applications, Third Edition, 2023.
\newblock \href {https://doi.org/10.1016/C2009-0-63921-0} {\path{doi:10.1016/C2009-0-63921-0}}.

\bibitem{Morris1994}
J.~R. Morris, C.~Z. Wang, K.~M. Ho, C.~T. Chan, Melting line of aluminum from simulations of coexisting phases, Physical Review B 49 (1994).
\newblock \href {https://doi.org/10.1103/PhysRevB.49.3109} {\path{doi:10.1103/PhysRevB.49.3109}}.

\bibitem{Sadigh2012}
B.~Sadigh, P.~Erhart, A.~Stukowski, A.~Caro, E.~Martinez, L.~Zepeda-Ruiz, Scalable parallel monte carlo algorithm for atomistic simulations of precipitation in alloys, Physical Review B - Condensed Matter and Materials Physics 85 (5 2012).
\newblock \href {https://doi.org/10.1103/PhysRevB.85.184203} {\path{doi:10.1103/PhysRevB.85.184203}}.

\bibitem{Li2024}
Z.~Li, S.~Scandolo, Efficient determination of free energies of non-ideal solid solutions via hybrid monte carlo simulations, Computer Physics Communications 304 (11 2024).
\newblock \href {https://doi.org/10.1016/j.cpc.2024.109307} {\path{doi:10.1016/j.cpc.2024.109307}}.

\bibitem{Plimpton1995}
S.~Plimpton, \href{https://www.sciencedirect.com/science/article/pii/S002199918571039X}{Fast parallel algorithms for short-range molecular dynamics}, Journal of Computational Physics 117~(1) (1995) 1--19.
\newblock \href {https://doi.org/10.1006/jcph.1995.1039} {\path{doi:10.1006/jcph.1995.1039}}.
\newline\urlprefix\url{https://www.sciencedirect.com/science/article/pii/S002199918571039X}

\bibitem{Thompson2022}
A.~P. Thompson, H.~M. Aktulga, R.~Berger, D.~S. Bolintineanu, W.~M. Brown, P.~S. Crozier, P.~J. in~'t Veld, A.~Kohlmeyer, S.~G. Moore, T.~D. Nguyen, R.~Shan, M.~J. Stevens, J.~Tranchida, C.~Trott, S.~J. Plimpton, \href{https://doi.org/10.1016/j.cpc.2021.108171}{Lammps - a flexible simulation tool for particle-based materials modeling at the atomic, meso, and continuum scales}, Computer Physics Communications 271 (2022) 108171.
\newblock \href {https://doi.org/10.1016/j.cpc.2021.108171} {\path{doi:10.1016/j.cpc.2021.108171}}.
\newline\urlprefix\url{https://doi.org/10.1016/j.cpc.2021.108171}

\bibitem{Babuji2019}
Y.~Babuji, A.~Woodard, Z.~Li, D.~S. Katz, B.~Clifford, R.~Kumar, L.~Lacinski, R.~Chard, J.~Wozniak, I.~Foster, M.~Wilde, K.~Chard, Parsl: Pervasive parallel programming in python, in: Proceedings of the 28th International Symposium on High-Performance Parallel and Distributed Computing, HPDC '19, Association, 2019, pp. 25--34.

\bibitem{Menon2021}
S.~Menon, Y.~Lysogorskiy, J.~Rogal, R.~Drautz, Automated free-energy calculation from atomistic simulations, Physical Review Materials 5 (10 2021).
\newblock \href {https://doi.org/10.1103/PhysRevMaterials.5.103801} {\path{doi:10.1103/PhysRevMaterials.5.103801}}.

\bibitem{Konning2025}
M.~D. Koning, Optimizing the driving function for nonequilibrium free-energy calculations in the linear regime: A variational approach, Journal of Chemical Physics 122 (2005).
\newblock \href {https://doi.org/10.1063/1.1860556} {\path{doi:10.1063/1.1860556}}.

\bibitem{Frenkel1984}
D.~Frenkel, A.~J. Ladd, New monte carlo method to compute the free energy of arbitrary solids. application to the fcc and hcp phases of hard spheres, The Journal of Chemical Physics 81 (1984).
\newblock \href {https://doi.org/10.1063/1.448024} {\path{doi:10.1063/1.448024}}.

\bibitem{Freitas2016}
R.~Freitas, M.~Asta, M.~D. Koning, Nonequilibrium free-energy calculation of solids using lammps, Computational Materials Science 112 (2016) 333--341.
\newblock \href {https://doi.org/10.1016/j.commatsci.2015.10.050} {\path{doi:10.1016/j.commatsci.2015.10.050}}.

\bibitem{Daw1984}
M.~S. Daw, M.~I. Baskes, Embedded-atom method: Derivation and application to impurities, surfaces, and other defects in metals, Physical Review B 29 (1984).
\newblock \href {https://doi.org/10.1103/PhysRevB.29.6443} {\path{doi:10.1103/PhysRevB.29.6443}}.

\bibitem{Finnis1984}
M.~W. Finnis, J.~E. Sinclair, A simple empirical n-body potential for transition metals, Philosophical Magazine A: Physics of Condensed Matter, Structure, Defects and Mechanical Properties 50 (1984).
\newblock \href {https://doi.org/10.1080/01418618408244210} {\path{doi:10.1080/01418618408244210}}.

\bibitem{Mendelev2009}
M.~I. Mendelev, M.~J. Kramer, R.~T. Ott, D.~J. Sordelet, D.~Yagodin, P.~Popel, Development of suitable interatomic potentials for simulation of liquid and amorphous cu-zr alloys, Philosophical Magazine 89 (2009).
\newblock \href {https://doi.org/10.1080/14786430902832773} {\path{doi:10.1080/14786430902832773}}.

\bibitem{Tang2012}
C.~Tang, P.~Harrowell, Predicting the solid state phase diagram for glass-forming alloys of copper and zirconium, Journal of Physics Condensed Matter 24 (6 2012).
\newblock \href {https://doi.org/10.1088/0953-8984/24/24/245102} {\path{doi:10.1088/0953-8984/24/24/245102}}.

\bibitem{Tang2025}
L.~Tang, W.~Xia, G.~Viswanathan, E.~Soto, K.~Kovnir, C.~Wang, Developing a neural network machine learning interatomic potential for molecular dynamics simulations of la–si–p systems, The Journal of Chemical Physics 163~(8) (2025) 084109.
\newblock \href {https://doi.org/10.1063/5.0284672} {\path{doi:10.1063/5.0284672}}.

\bibitem{Sun2023}
Y.~Sun, M.~I. Mendelev, F.~Zhang, X.~Liu, B.~Da, C.~Z. Wang, R.~M. Wentzcovitch, K.~M. Ho, Ab initio melting temperatures of bcc and hcp iron under the earth’s inner core condition, Geophysical Research Letters 50 (3 2023).
\newblock \href {https://doi.org/10.1029/2022GL102447} {\path{doi:10.1029/2022GL102447}}.

\bibitem{Mendelev2015}
M.~I. Mendelev, F.~Zhang, Z.~Ye, Y.~Sun, M.~C. Nguyen, S.~R. Wilson, C.~Z. Wang, K.~M. Ho, Development of interatomic potentials appropriate for simulation of devitrification of al90sm10 alloy, Modelling and Simulation in Materials Science and Engineering 23 (2015).
\newblock \href {https://doi.org/10.1088/0965-0393/23/4/045013} {\path{doi:10.1088/0965-0393/23/4/045013}}.

\bibitem{Tang2020}
L.~Tang, Z.~J. Yang, T.~Q. Wen, K.~M. Ho, M.~J. Kramer, C.~Z. Wang, Development of interatomic potential for al-tb alloys using a deep neural network learning method, Physical Chemistry Chemical Physics 22 (2020).
\newblock \href {https://doi.org/10.1039/d0cp01689f} {\path{doi:10.1039/d0cp01689f}}.

\bibitem{Mendelev2019}
M.~Mendelev, Y.~Sun, F.~Zhang, C.~Wang, K.~Ho, Development of a semi-empirical potential suitable for molecular dynamics simulation of vitrification in cu-zr alloys, Journal of Chemical Physics 151 (2019).
\newblock \href {https://doi.org/10.1063/1.5131500} {\path{doi:10.1063/1.5131500}}.

\bibitem{Abramo2015}
M.~C. Abramo, C.~Caccamo, D.~Costa, P.~V. Giaquinta, G.~Malescio, G.~Munaò, S.~Prestipino, On the determination of phase boundaries via thermodynamic integration across coexistence regions, Journal of Chemical Physics 142 (2015).
\newblock \href {https://doi.org/10.1063/1.4921884} {\path{doi:10.1063/1.4921884}}.

\bibitem{Leite2016}
R.~P. Leite, R.~Freitas, R.~Azevedo, M.~D. Koning, The uhlenbeck-ford model: Exact virial coefficients and application as a reference system in fluid-phase free-energy calculations, Journal of Chemical Physics 145 (11 2016).
\newblock \href {https://doi.org/10.1063/1.4967775} {\path{doi:10.1063/1.4967775}}.

\bibitem{Leite2019}
R.~P. Leite, M.~de~Koning, Nonequilibrium free-energy calculations of fluids using lammps, Computational Materials Science 159 (2019) 316--326.
\newblock \href {https://doi.org/10.1016/j.commatsci.2018.12.029} {\path{doi:10.1016/j.commatsci.2018.12.029}}.

\bibitem{Yang2018}
L.~Yang, Y.~Sun, Z.~Ye, F.~Zhang, M.~I. Mendelev, C.~Z. Wang, K.~M. Ho, A self-contained algorithm for determination of solid-liquid equilibria in an alloy system, Computational Materials Science 150 (2018) 353--357.
\newblock \href {https://doi.org/10.1016/j.commatsci.2018.04.028} {\path{doi:10.1016/j.commatsci.2018.04.028}}.

\bibitem{Wilson2015}
S.~R. Wilson, K.~G. Gunawardana, M.~I. Mendelev, Solid-liquid interface free energies of pure bcc metals and b2 phases, Journal of Chemical Physics 142 (2015).
\newblock \href {https://doi.org/10.1063/1.4916741} {\path{doi:10.1063/1.4916741}}.

\bibitem{Pedersen2018}
U.~R. Pedersen, T.~B. Schrøder, J.~C. Dyre, Phase diagram of kob-andersen-type binary lennard-jones mixtures, Physical Review Letters 120 (4 2018).
\newblock \href {https://doi.org/10.1103/PhysRevLett.120.165501} {\path{doi:10.1103/PhysRevLett.120.165501}}.

\bibitem{Andersson2002}
J.~O. Andersson, T.~Helander, L.~Höglund, P.~Shi, B.~Sundman, Thermo-calc \& dictra, computational tools for materials science, Calphad: Computer Coupling of Phase Diagrams and Thermochemistry 26 (2002).
\newblock \href {https://doi.org/10.1016/S0364-5916(02)00037-8} {\path{doi:10.1016/S0364-5916(02)00037-8}}.

\bibitem{Otis2017}
R.~Otis, Z.-K. Liu, \href{https://openresearchsoftware.metajnl.com/articles/10.5334/jors.140}{pycalphad: Calphad-based computational thermodynamics in python}, Journal of Open Research Software 5~(1) (2017) 1.
\newblock \href {https://doi.org/10.5334/jors.140} {\path{doi:10.5334/jors.140}}.
\newline\urlprefix\url{https://openresearchsoftware.metajnl.com/articles/10.5334/jors.140}

\bibitem{Okamoto2012}
H.~Okamoto, Cu-zr (copper-zirconium) (10 2012).
\newblock \href {https://doi.org/10.1007/s11669-012-0077-1} {\path{doi:10.1007/s11669-012-0077-1}}.

\bibitem{Sander2017}
D.~Sander, S.~O. Valenzuela, D.~Makarov, C.~H. Marrows, E.~E. Fullerton, P.~Fischer, J.~McCord, P.~Vavassori, S.~Mangin, P.~Pirro, B.~Hillebrands, A.~D. Kent, T.~Jungwirth, O.~Gutfleisch, C.~G. Kim, A.~Berger, \href{https://doi.org/10.1088/1361-648X/aa680e}{The 2017 magnetism roadmap}, Journal of Physics D: Applied Physics 50~(36) (2017) 363001.
\newblock \href {https://doi.org/10.1088/1361-648X/aa680e} {\path{doi:10.1088/1361-648X/aa680e}}.
\newline\urlprefix\url{https://doi.org/10.1088/1361-648X/aa680e}

\bibitem{pycalphad2025}
{PyCALPHAD: Computational Thermodynamics Documentation}, \url{https://pycalphad.org/docs/latest/}, accessed: 2025-09-18 (2025).

\end{thebibliography}

\end{document}